\documentclass[sigconf]{acmart}

\AtBeginDocument{%
  \providecommand\BibTeX{{%
    \normalfont B\kern-0.5em{\scshape i\kern-0.25em b}\kern-0.8em\TeX}}}
    
\copyrightyear{2023}
\acmYear{2023}
\setcopyright{acmlicensed}\acmConference[GECCO '23]{Genetic and Evolutionary Computation Conference}{July 15--19, 2023}{Lisbon, Portugal}
\acmBooktitle{Genetic and Evolutionary Computation Conference (GECCO '23), July 15--19, 2023, Lisbon, Portugal}
\acmPrice{15.00}
\acmDOI{10.1145/3583131.3590520}
\acmISBN{979-8-4007-0119-1/23/07}

\begin{document}

\title{Universal Mechanical Polycomputation in Granular Matter}

\author{Atoosa Parsa}
\affiliation{%
  \institution{University of Vermont}
  \city{Burlington}
  \state{Vermont}
  \country{USA}
}
\email{atoosa.parsa@uvm.edu}

\author{Sven Witthaus}
\affiliation{%
  \institution{Yale University}
  \city{New Haven}
  \state{Connecticut}
  \country{USA}
}
\email{sven.witthaus@yale.edu}

\author{Nidhi Pashine}
\affiliation{%
  \institution{Yale University}
  \city{New Haven}
  \state{Connecticut}
  \country{USA}
}
\email{nidhi.pashine@yale.edu}

\author{Corey S. O'Hern}
\affiliation{%
  \institution{Yale University}
  \city{New Haven}
  \state{Connecticut}
  \country{USA}
}
\email{corey.ohern@yale.edu}

\author{Rebecca Kramer-Bottiglio}
\affiliation{%
  \institution{Yale University}
  \city{New Haven}
  \state{Connecticut}
  \country{USA}
}
\email{rebecca.kramer@yale.edu}

\author{Josh Bongard}
\affiliation{%
  \institution{University of Vermont}
  \city{Burlington}
  \state{Vermont}
  \country{USA}
}
\email{josh.bongard@uvm.edu}

\renewcommand{\shortauthors}{Parsa, et al.}


\begin{abstract}
Unconventional computing devices are increasingly of interest as they can operate in environments hostile to silicon-based electronics, or compute in ways that traditional electronics cannot. Mechanical computers, wherein information processing is a material property emerging from the interaction of components with the environment, are one such class of devices. This information processing can be manifested in various physical substrates, one of which is granular matter. In a granular assembly, vibration can be treated as the information-bearing mode. This can be exploited to realize ``polycomputing'': materials can be evolved such that a single grain within them can report the result of multiple logical operations simultaneously at different frequencies, without recourse to quantum effects. Here, we demonstrate the evolution of a material in which one grain acts simultaneously as two different NAND gates at two different frequencies. NAND gates are of interest as any logical operations can be built from them. Moreover, they are nonlinear thus demonstrating a step toward general-purpose, computationally dense mechanical computers. Polycomputation was found to be distributed across each evolved material, suggesting the material’s robustness. With recent advances in material sciences, hardware realization of these materials may eventually provide devices that challenge the computational density of traditional computers.
\end{abstract}



\maketitle


\section{Introduction}

Traditionally, combinational logic operations are embedded in silicon based electronic devices. However, in light of the recent advances in chemical, physical and material sciences, unconventional practices of information processing in engineered materials are now being explored \cite{yasuda2021mechanical}. Quantum computing \cite{hey1999quantum}, DNA computing \cite{amos2002topics}, neuromorphic computing \cite{markovic2020physics}, optical computing \cite{solli2015analog}, and physical reservoir computing \cite{nakajima2020physical} are all examples of the new forms of embedding computation. Moreover, attempts have been made to overcome the limitations of the conventional Integrated Circuits via combining them with soft conductive materials to develop kinematically reconfigurable electrical circuits \cite{el2022mechanical}. Exploring these new computational paradigms provides novel opportunities for the development of advanced forms of intelligent matter in which sensing, decision making and actuation are combined \cite{kaspar2021rise}. One recent example is logic-enabled textiles that can be embedded into robotic wearables \cite{rajappan2022logic}.

\begin{figure}[h!]
    \centering
    \includegraphics[width=\linewidth]{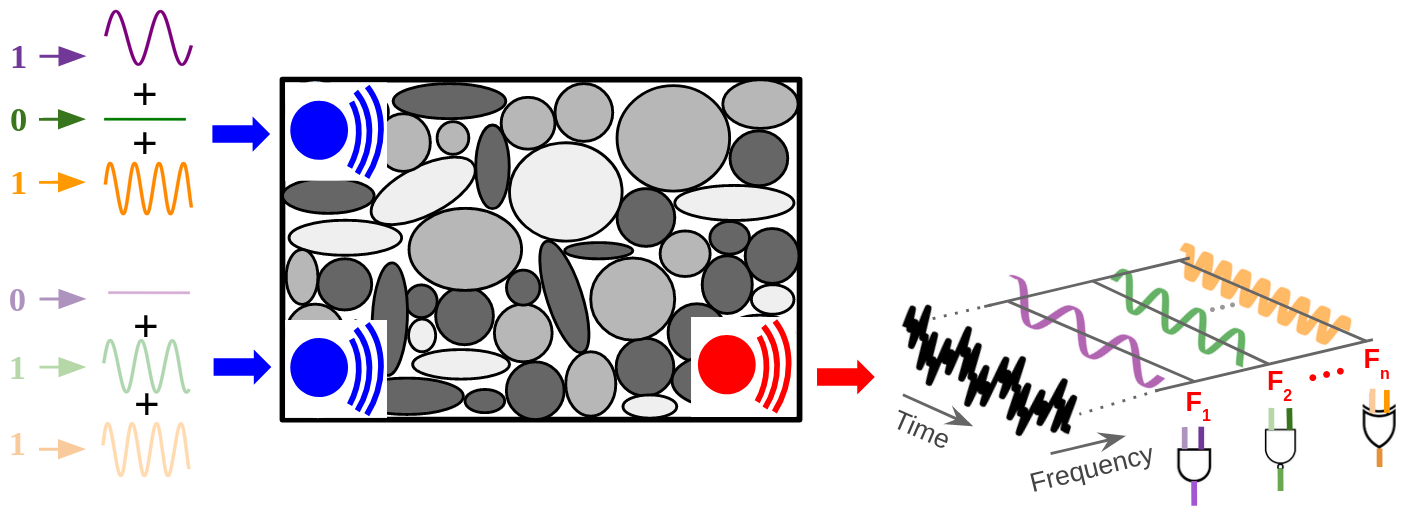}
    \caption{The size, shape, and position of the grains in a granular metamaterial can be evolved such that it ``polycomputes'': different logical functions are computed at the same place, at the same time, in the material. This can be accomplished if bits, encoded as vibrations, are combined by summing vibrations at different frequencies. The power of response at a different output grain (red), at those frequencies, can be interpreted as the output of the computation performed on those different bit pairs.}\label{fig1}
\end{figure}

Logical functions embedded in metamaterials have also been explored to some extent. Non-intuitive exotic properties of these engineered materials make them great candidates as physical substrates to perform computation \cite{silva2014performing}. In \cite{ion2017digital} authors have proposed 3D printed cells with embedded bistable springs that implement simple logic functions using mechanical signals. In \cite{treml2018origami} origami units have been designed that can be used to develop a framework for embedding computation into the robotic structures. \cite{serra2019turing} has exploited a nonlinear mass-spring-damper model to design a universal logic gate.

Recently, the granular metamaterials \cite{kim2019wave} are gaining special interest as the physical substrates for computation. Here, the mechanical vibrations can be used as the information-bearing mode to realize various logical functions in a granular matter (Fig.~\ref{fig1}). Because these materials are made of individual particles with diverse material properties, they can also be responsive to external stimuli (like temperature or electromagnetic fields) and thus possess a higher potential for an increased computational density. In \cite{Feng14} OR and AND logical gates are designed in a granular chain of chrome steel grains. In \cite{bilal2017bistable} the bistability of an array of spiral-spring resonators is exploited to build transistor-like switches. In all of the aforementioned works (except \cite{ourPaper}), the computational modules are hand-designed by human experts. As we move to higher-order parameter spaces and explore more complicated structures, the need for an automated optimization pipeline to design computational metamaterials seems unavoidable. Moreover, in none of the previous works the scalability of the designed units to realize the universal computational gates is explored. 

In this paper, we first use Evolutionary Algorithms to design granular assemblies that can compute a NAND function. Next, by leveraging the superposition principle in the frequency response of the coupled granular system under external vibrations, we explore ``polycomputation'' in the material and provide the optimization results showing NAND functionality computed at two different frequencies at the same time and in the same place. We provide an exploratory analysis of the distribution of the computation throughout the material and then discuss a possible hardware platform for the physical realization of our designed structures. 

\section{Generalization}
Boolean logic gates are at the root of any computational model. Logical NAND gates are of particular interest because any Boolean function can be implemented using circuits constructed from only NAND gates. Therefore, to take a step toward general-purpose computation embedded in granular metamaterials, in this paper we focus on evolving NAND gates. In this section, we start with introducing the digital abstraction in our physical platform. We then describe the details of the granular material simulator used by the Evolutionary Algorithm. After discussing the EA itself, we present the results of evolving a NAND gate in a material and then show common patterns that consistently arise across independent solutions, suggesting deeper insight into how various materials can realize new forms of computation.

\subsection{Computation in Granular Matter}
\begin{figure}[h!]
    \centering
    \includegraphics[width=\linewidth]{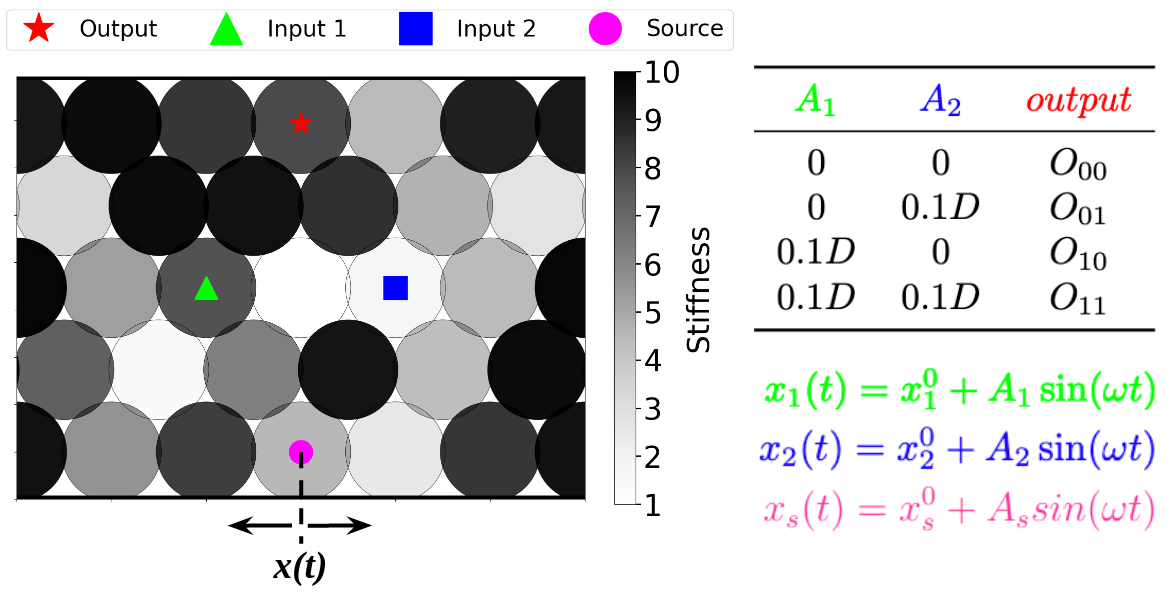}
    \caption{System setup. A logic gate with two inputs (green and blue particles), one output (red particle) and one power source (magenta particle) embedded in a 2D granular assembly with particles of various stiffness properties (indicated as different shades of gray). $A_1$ and $A_2$ are the amplitudes of oscillations applied in $x$ direction to the input ports which will result in displacements from the initial positions ($x^0_1$ and $x^0_2$). $A_s$ is the amplitude of the source signal. $\omega$ is the operational frequency. The truth table is shown on the right: $D$ represents the diameter of particles. $O_{ij} (i,j \in {0, 1})$ is the magnitude of the vibrations in the output.}\label{fig2}
\end{figure}

To encode information as sequences of bits, we must first establish a useful representation based on our physical substrate. In digital electronic systems, inputs and outputs are electrical voltages. In our analogous mechanical system, inputs and outputs are vibrations. The material is a two-dimensional assembly of circular particles placed on a triangular lattice (Fig.~\ref{fig2}). The macroscopic behavior of such a granular system is determined by the properties of individual particles such as their arrangement, mass, modulus, and shape. In the present work, the EA was only allowed to vary a particle's stiffness ratio: it could be set to any value in $[1, 10]$.  These microscopic properties influence how vibrations propagate through the material; our goal here is to evolve particle stiffnesses such that the material's vibrational response resembles the operation of a desired logic gate.

In order to apply the input signal, two particles are arbitrarily chosen as input ports (particles with green and blue markers in Fig.~\ref{fig2}). The red particle represents the output port used to observe the system's response. A fourth particle (magenta) is connected to an external power source, which is realized as constant vibration of that particle, regardless of the vibrations supplied to the inputs in the material. All input and power vibrations are sinusoidal waves with amplitude $A_i$ and frequency $\omega$ applied in the $x$ direction.

Other particles may vibrate more or less in $x$ and $y$ directions. Applying this vibration to the particle causes a displacement from its initial position ($x_i^0$). The amplitude of this displacement signal over time ($x_i(t)$) is interpreted as a bit: (considering the amplitude of the power source as a baseline) small or no displacement is interpreted as a $`0`$; high magnitude of displacement is interpreted as a $`1`$. The truth table in Fig.~\ref{fig2} shows this representation. Oscillation amplitude arriving at the output port ($O_ij$) at a pre-determined frequency is recorded in each of the four input cases ($`00`$, $`01`$, $`10`$, and $`11`$). This quantity is then used to calculate the system's gain ($G_{ij}$) defined as:

\begin{equation}\label{eq1}
G_{ij}(\omega) = \frac{\hat{f}(O_{ij})}{\hat{f}({\mathrm{in}}_i)+\hat{f}({\mathrm{in}}_j)} \quad i,j \in{0, 1}
\end{equation}

In this equation, $\hat{f}$ is the fast Fourier transform at the driving frequency ($\omega$), which is the frequency applied to the input and power ports. For the material to resemble a NAND gate, we expect to see high amplitude oscillation at the output in the $`00`$, $`01`$ and $`10`$ cases (which translates to high $G_{00}$, $G_{01}$ and $G_{10}$), but a low amplitude in the $`11`$ case (low $G_{11}$). Using this formulation, we can evaluate each candidate granular assembly of particles and determine its similarity to the desired logical function. To select for materials that are increasingly NAND-like, the fitness function (which was minimized during the optimization) was defined as follows:

\begin{equation}\label{eq2}
    F = \max{(|1-G_{00}(\omega)|, |1-G_{01}(\omega)|, |1-G_{10}(\omega)|, |0-G_{11}(\omega)|)}
\end{equation}

\subsection{Simulation}
We simulated our granular system using the Discrete Element Method (DEM) \cite{wu2019active}. The particles are assumed to be frictionless circular disks with a fixed diameter ($D$), placed on a $5$ by $6$ two-dimensional hexagonal lattice (a total number of $30$ particles) with periodic boundary in the $x$ direction. Gravity is ignored and the particle-particle interaction is modeled as a purely repulsive force with a Lennard-Jones potential. The Fast Inertial Relaxation Engine (FIRE) \cite{bitzek2006structural} is used to adjust the initial positions of the particles and ensure a statistically stable particle packing at the start of the simulation. Table~\ref{table1} reports the main parameters used in our experiments.  

\begin{table}[htb!]
\caption{Experimental Setup.}
\centering
\begin{tabular}{cc|cc} 
\hline
\multicolumn{2}{c}{\bfseries Simulator} & \multicolumn{2}{c}{\bfseries Optimizer} \\
 \hline\hline
    \textit{Particle Diameter} & $0.1$ & \textit{Algorithm} & AFPO\\
    \textit{Particle Mass} & $1$ & \textit{Population Size} & $50$\\
    \textit{Stiffness Ratio} & $\in [1, 10]$ & \textit{Generations} & $500$\\
    \textit{Packing Fraction} & $0.91$ &  \textit{Runs} & $5$\\
    \textit{Simulation Time} & $1e4$ & \textit{Mutation} & Gaussian \\
\hline
\end{tabular}
\label{table1}
\end{table}

Source code for the experiments in this paper can be found at: \href{https://github.com/AtoosaParsa/gecco-2023}{https://github.com/AtoosaParsa/gecco-2023}.

\subsection{Optimization}
There are many parameters in a granular system that affect its macroscopic response to the applied vibrations: particle shapes, sizes, placements and material properties. These affect the normal modes of the system, and consequently the propagation or suppression of the input and power vibrations. Deciding on the optimal parameters to achieve a desired behavior in the material is a challenging task for a human expert and thus optimization algorithms can be used to automate this design process. Evolutionary algorithms have long been used effectively for many real-world applications; more recently they have been applied for designing granular matter \cite{miskin2013adapting} and computational granular matter \cite{parsa2022evolving}. In this section, we used the Age-Fitness Pareto Optimization (AFPO) algorithm \cite{schmidt2010age} to find granular configurations that act as a single NAND gate at a pre-determined frequency.

\begin{figure}[h!]
    \centering
    \includegraphics[width=\linewidth]{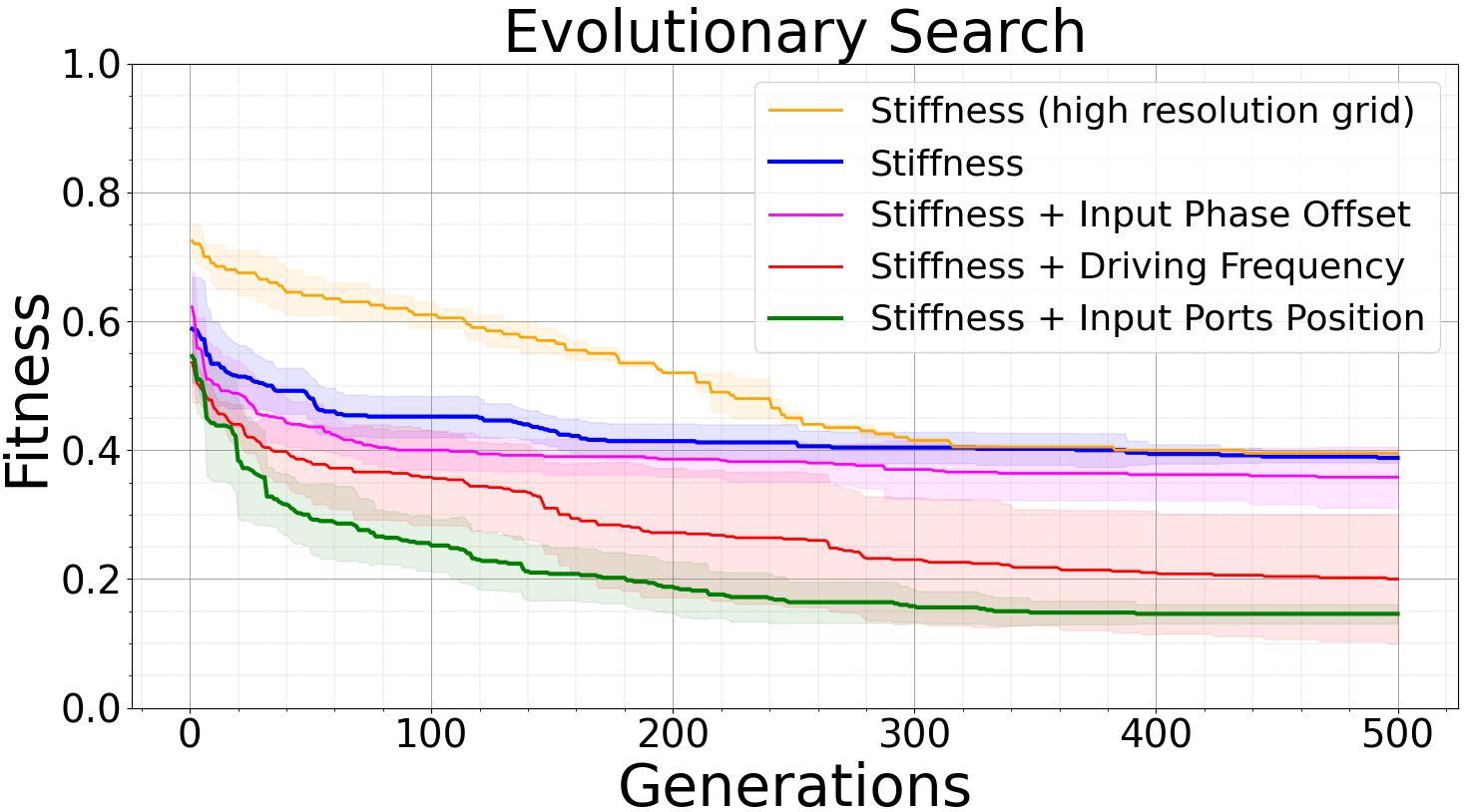}
    \caption{Examination of the parameter space. The evolutionary progress of $5$ experiments, each with a different set of optimized parameters (details in the text) is plotted. The plots show the average fitness of the population per generation, averaged over $5$ independent runs. The orange plot indicates the experiment with a $10 \times 10$ grid of particles. Because of the limited computational resources, only $2$ evolutionary runs were performed in this case.}\label{fig3}
\end{figure}

Each individual in the population is a granular assembly made of particles with different stiffness values. There are many design parameters in this system that can be included in the optimization. In initial experiments, it was found that the positions of the ports can also affect the gate's operation. In these experiments, we examined $5$ setups with different parameter sets including the stiffness vector, frequency of the input vibrations, phase offset between the input vibrations, and placement of the input ports (Fig.\ref{fig3}). We cannot conclude that one algorithm variant had statistically significantly better performance than another, but the results does suggest that input port position can be helpful in the optimization. For this reason, we expanded the evolutionary algorithm to also evolve the position of the input ports. To reduce the dimensionality of the search space, we fixed the position of the source and output ports. We used a direct encoding, with a genome that is made of two parts. The first part is a real vector of length $30$ (the lattice is $5 \times 6$) that represents the stiffness of each particle (the values range between $1$ (soft) and $10$ (stiff)). The second part of the genome contains two integers that dictate the input port positions. Two mutation operators were employed: the first acts on the real vector and is a Gaussian mutation operator (with standard deviation of $0.1$). The second randomizes the position of one of the two input ports, chosen randomly. We performed $5$ independent trials, each with a different random initial population.

\subsection{Results}
\begin{figure}[h!]
    \centering
    \includegraphics[scale=0.31]{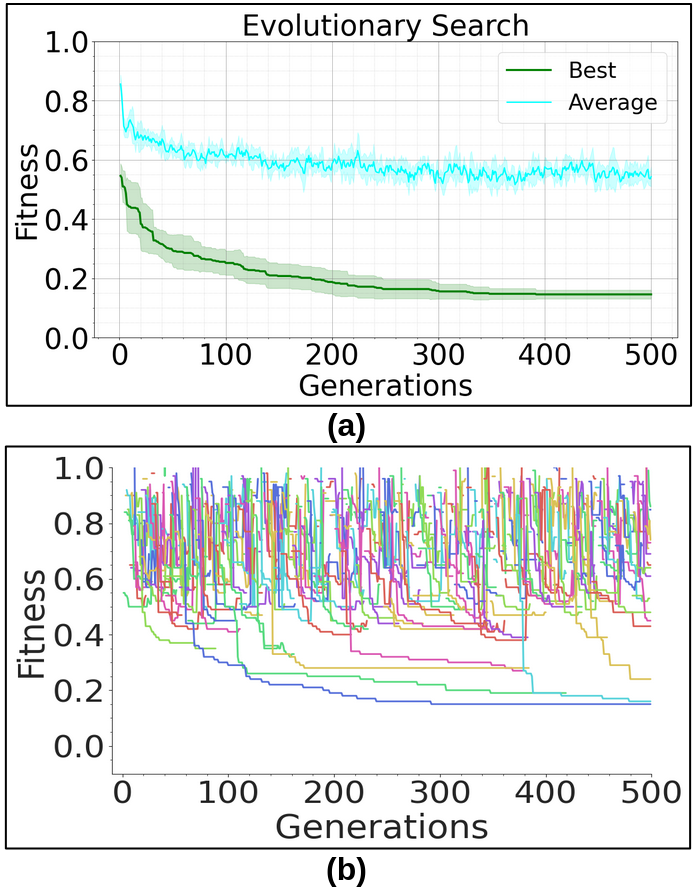}
    \caption{(a) Evolutionary progress across the five trials dedicated to evolving a single NAND gate into the material. (b) reports evolutionary change within one trial. Each line reports the best individual within that AFPO clade at each generation. Terminated lines denote clade extinction events.}\label{fig4}
\end{figure}

\begin{figure}[h!]
    \centering
    \includegraphics[width=\linewidth]{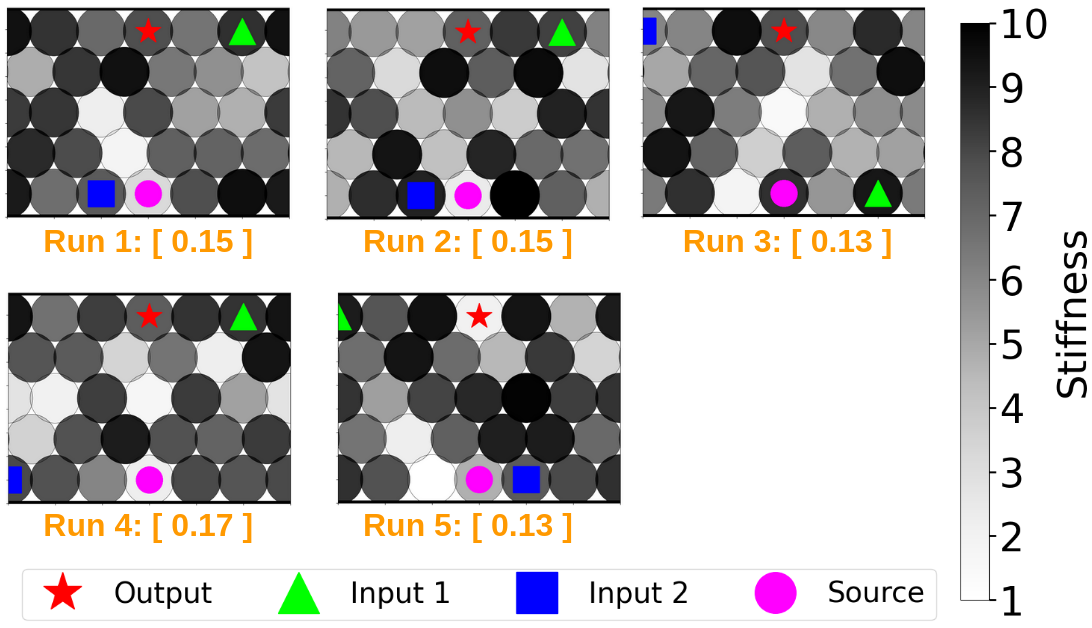}
    \caption{The five best materials drawn from the five single-NAND evolutionary trials. Fitness (Eqn.~\ref{eq2}) is reported below each material.}\label{fig5}
\end{figure}

Fig.~\ref{fig4} shows the evolutionary progress. Panel (a) shows the fitness (Eqn.~\ref{eq1}) decreasing over generations and reaching a value of $~0.1$ at the end of the optimization. The best solutions from each of the independent evolutionary runs are shown in Fig.~\ref{fig5}. 

\begin{figure}[h!]
    \centering
    \includegraphics[scale=0.37]{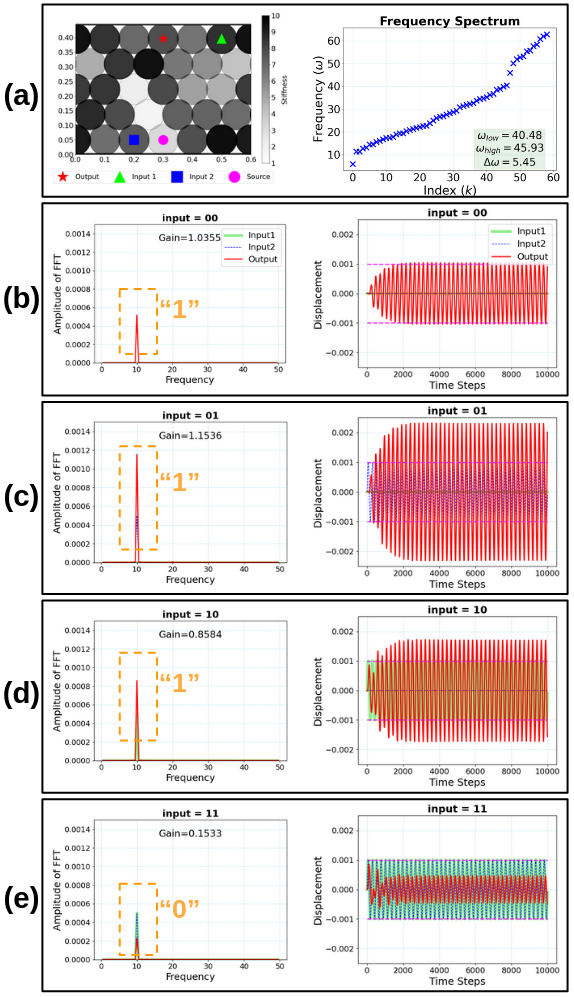}
    \caption{One of the five best materials acting as a single NAND gate at $\omega=10 Hz$. (a): particle configuration (left); frequency spectrum (right). (b-e): the response of the input (blue and green) and output (red) particles in the frequency (left) and time (right) domains.}\label{fig6}
\end{figure}

We verify the performance of one of the evolved solutions by testing the configuration's response when different vibrations are applied to the input ports. Fig.~\ref{fig6}b-e shows the response in frequency and time space when supplying the four different input signals. We see a low amplitude of oscillation when both of the input ports are activated. This confirms the material functions as a logical NAND gate at $\omega=10 Hz$.

\section{Distributed Computation}
The dynamics of a granular system is mathematically equivalent to a system of coupled oscillators. In such a system, the steady-state response to an arbitrary external oscillation is a superposition of the vibrational normal modes. The normal modes (or eigenfrequencies) of the granular assembly are determined by the material properties of the individual particles and the boundary conditions of the confining structure. So fundamentally, our proposed optimization platform is searching for a configuration that excites normal modes compatible with our desired logical function (which is the NAND operation at frequency $\omega=10 Hz$).

It can be informative to inspect the system's behavior at other frequencies in the spectrum and to investigate if there is any amount of ``NAND-ness'' present in the system at those frequencies, without any explicit selection pressure in the evolutionary optimization. Because the fitness function (Eqn.~\ref{eq2}) is designed to only consider the worst case of the four gain values, We define the following metric to combine the system's gains in all four cases and obtain one single number to measure the amount of ``NAND-ness'' in the material:

\begin{equation}\label{eq3}
    M_{\textrm{``NAND-ness''}}(\omega) = \frac{G_{00}(\omega) G_{01}(\omega) G_{10}(\omega)}{G_{11}(\omega)}
\end{equation}

\begin{figure}[h!]
    \centering
    \includegraphics[scale=0.32]{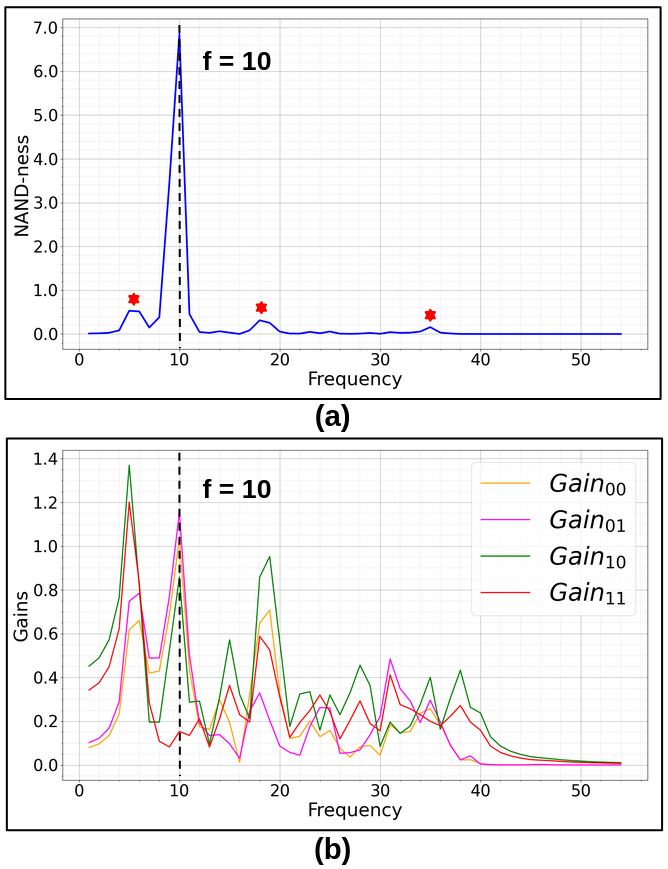}
    \caption{For one of the most fit materials evolved to exhibit NAND-ness at $\omega=10 Hz$, latent potential to act as a NAND gate at other frequencies can be seen in (a). Red markers show the potential frequencies (other than $10 Hz$) for acting as a NAND gate. (b) shows how the four NAND cases contribute or detract from that potential.}\label{fig7}
\end{figure}

Fig.~\ref{fig7} presents the result of sweeping through the frequency spectrum and calculating the ``NAND-ness'' at each frequency for one of the most fit materials with the evolved stiffness vector and port placements. As we expected, there is a high peak at $f=10 Hz$ which is the frequency used in the optimization. But we also notice that the plot is not completely flat at frequencies other than $10 Hz$: there are noticeable peaks at $~7 Hz$, $~18 Hz$, and $~35 Hz$ (marked with red stars in Fig.~\ref{fig7}.a). Investigating the gains (Fig.~\ref{fig7}.b) at these frequencies indicates some of the four cases comprising the NAND operation are being performed better or worse at these other frequencies. Taken together, this suggests that, with a multiobjective optimization method, we may be able to evolve NAND gates that operate simultaneously at different frequencies. This would in turn increase the computational density of the material. This idea is investigated in the next section.

\begin{figure*}[h!]
    \centering
    \includegraphics[scale=0.26]{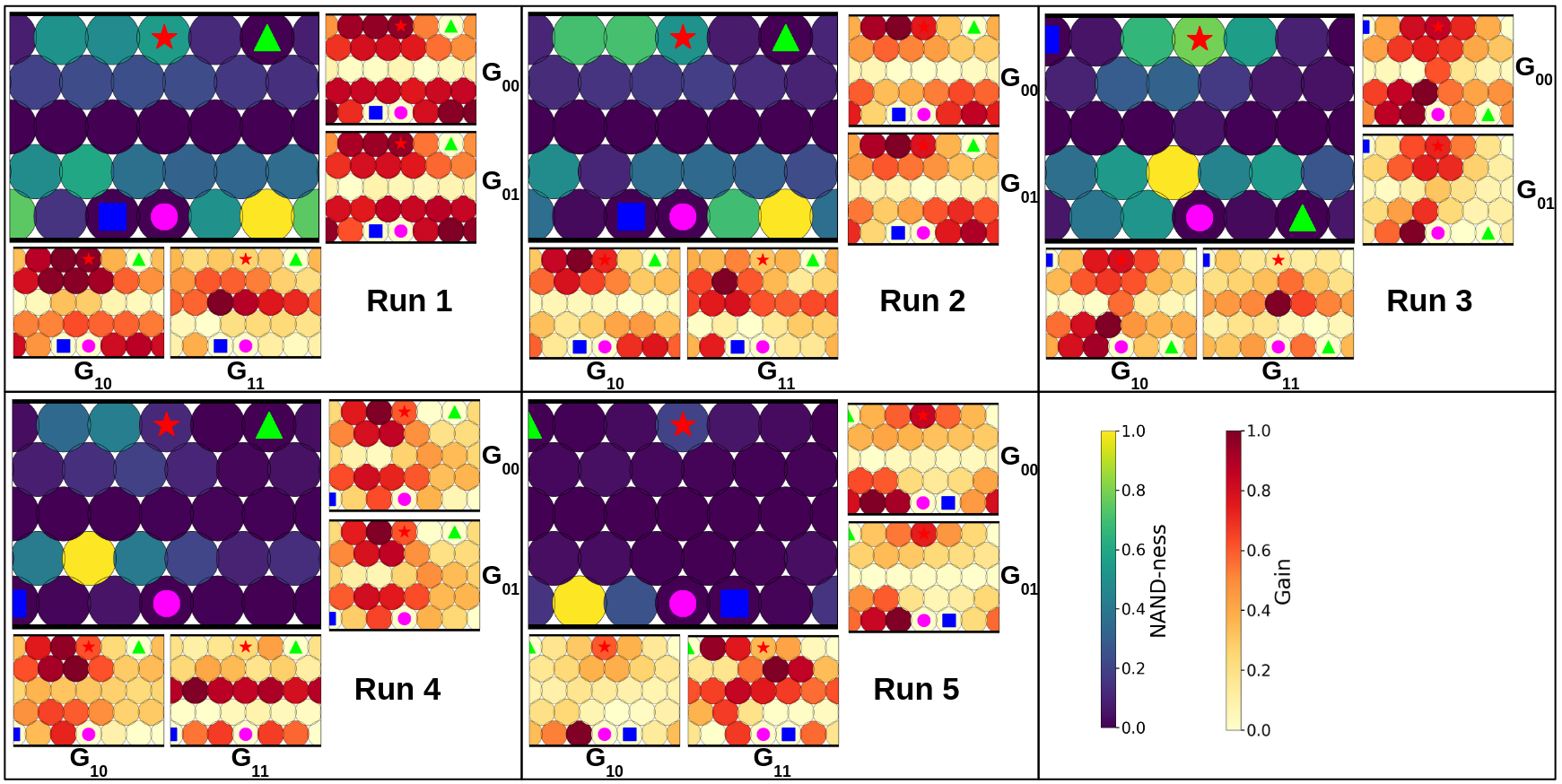}
    \caption{Distributed computation was found in the five best materials evolved to act as a single NAND gate. Lighter colors are better in the NAND-ness and $G_{11}$; darker is better in the $G_{00}$, $G_{01}$, and $G_{10}$ subplots}\label{fig8}
\end{figure*}

To further how these materials evolved to compute the NAND function, we looked at the amount of NAND-ness at every particle in the five best materials (Fig.~\ref{fig8}). Surprisingly, in all five cases, there was another particle that was much better at NAND operation than the pre-designated output particle (red star). This suggests that evolving output port position will, in future experiments, greatly facilitate the evolution of computation in these kinds of materials. 

\section{Polycomputation}
Silicon-based digital electronic computing devices have long stood as the state of the art for efficient, fast, and scalable computing \cite{hooker2021hardware}. However, logical components in Integrated Circuits are only capable of serial operation. This means that in order to implement more complex operations, many logic gates need to be connected in parallel or series and thus the number of required transistors will increase with the complexity of the computation. Moreover, wiring between transistors further limits the integration density. Polycomputing is defined as the ability of the same physical substrate to perform multiple functions in the same place and at the same time \cite{bongard2022there} and might eventually provide new paths to realize computationally dense materials. Granular metamaterials have been shown to possess this ability to simultaneously compute two logical functions (AND and XOR) by carrying out the operations at different frequencies of vibrations \cite{ourPaper}. In this section, this idea has been extended to NAND gates at multiple frequencies.

\begin{figure}[h!]
    \centering
    \includegraphics[scale=0.31]{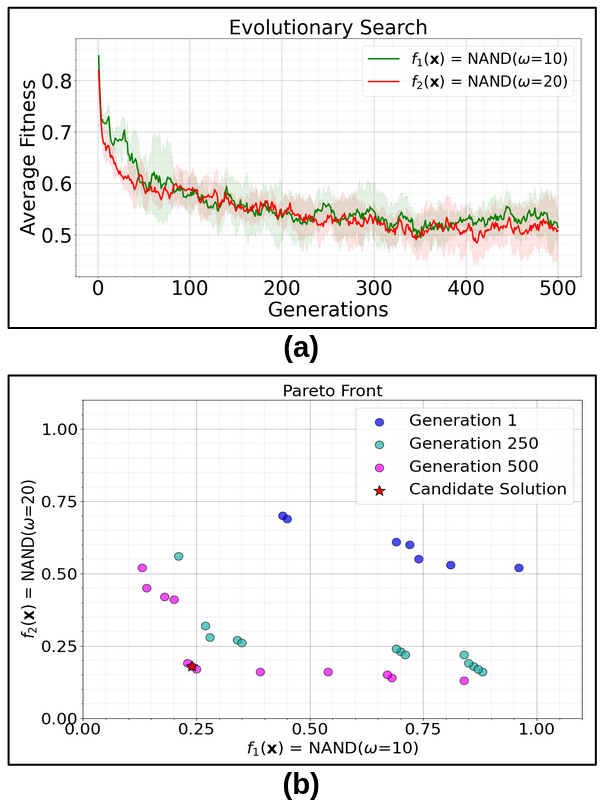}
    \caption{Evolution of polycomputation. The progress of each of the fitness functions during evolution is plotted in panel (a), these plots are averaged over $5$ evolutionary trials. Panel (b) shows the Pareto front for one of the trials, at three different stages during the optimization. The selected solution from the last generation is marked with a red star.}\label{fig10}
\end{figure}

The simulator and experimental setup are the same as described in the previous section. The only difference is the number of objectives in the evolutionary algorithm. Here, a tri-objective optimization method was employed: the AFPO age objective was minimized, as  was NAND-ness at $\omega=10 Hz$ ($f_1$) and NAND-ness at $\omega=20$ ($f_2$). We used the same fitness function defined in Eqn.~\ref{eq2}. Because of the increased number of objectives, we also increase the population size to $100$ to minimize the likelihood of premature convergence (however this may have occurred in some of the trials). Fig. \ref{fig10} presents the result of this optimization.

\begin{figure*}[h!]
    \centering
    \includegraphics[width=\textwidth]{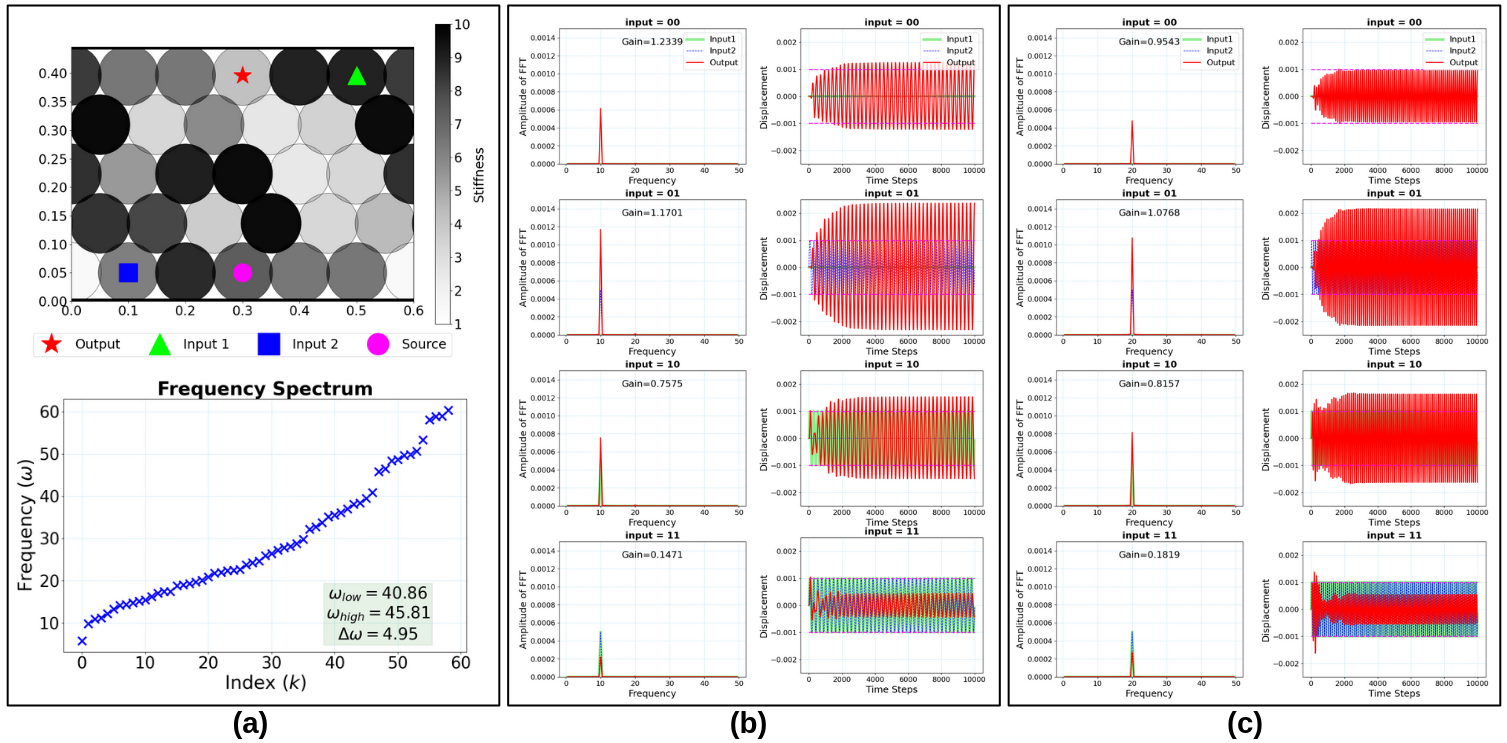}
    \caption{Polycomputation. An optimized configuration of particles functions as a NAND gate at two different frequencies: $\omega=10 Hz$ in panel (b) and $\omega=20 Hz$ in panel (c)}\label{fig11}
\end{figure*}

The material at the knee of the Pareto optimal set after $500$ generations was chosen for analysis (red marker in Fig.~\ref{fig10}b). It was evaluated at $\omega=10 Hz$ and $\omega=20$, and the material's response is plotted in the frequency and time domains in Fig.~\ref{fig11}.

\begin{figure}[h!]
    \centering
    \includegraphics[scale=0.21]{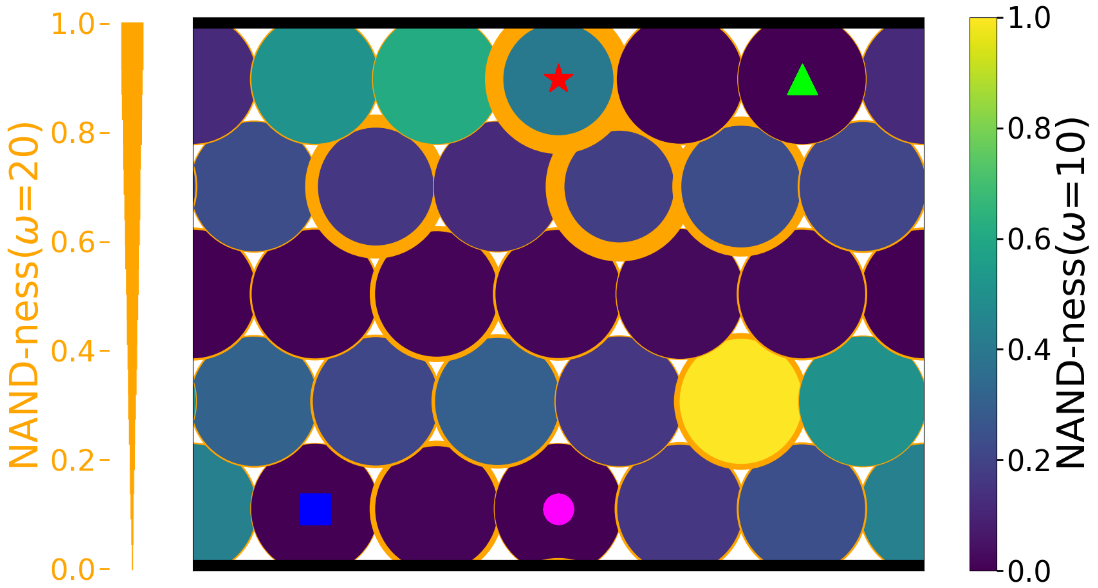}
    \caption{Distribution of polycomputation. Heatmap shows the amount of NAND-ness in each particle at $\omega=10 Hz$. Thickness of each particle's perimeter represents NAND-ness at $\omega=20 Hz$.}\label{fig12}
\end{figure}

Fig.~\ref{fig12} reports that the output particle has moderate amounts of NAND-ness at both low ($\omega=10 Hz$) and high ($\omega=20 Hz$) frequencies. Moreover, it also shows that other particles throughout the material have specialized to exhibit varying amounts of NAND-ness at the two frequencies. As in previous experiments, one particular particle exhibits as much, if not more NAND-ness, as the output particle, for both frequencies.

\section{Robustness}
\begin{figure*}[h!]
\centering
\includegraphics[scale=0.30]{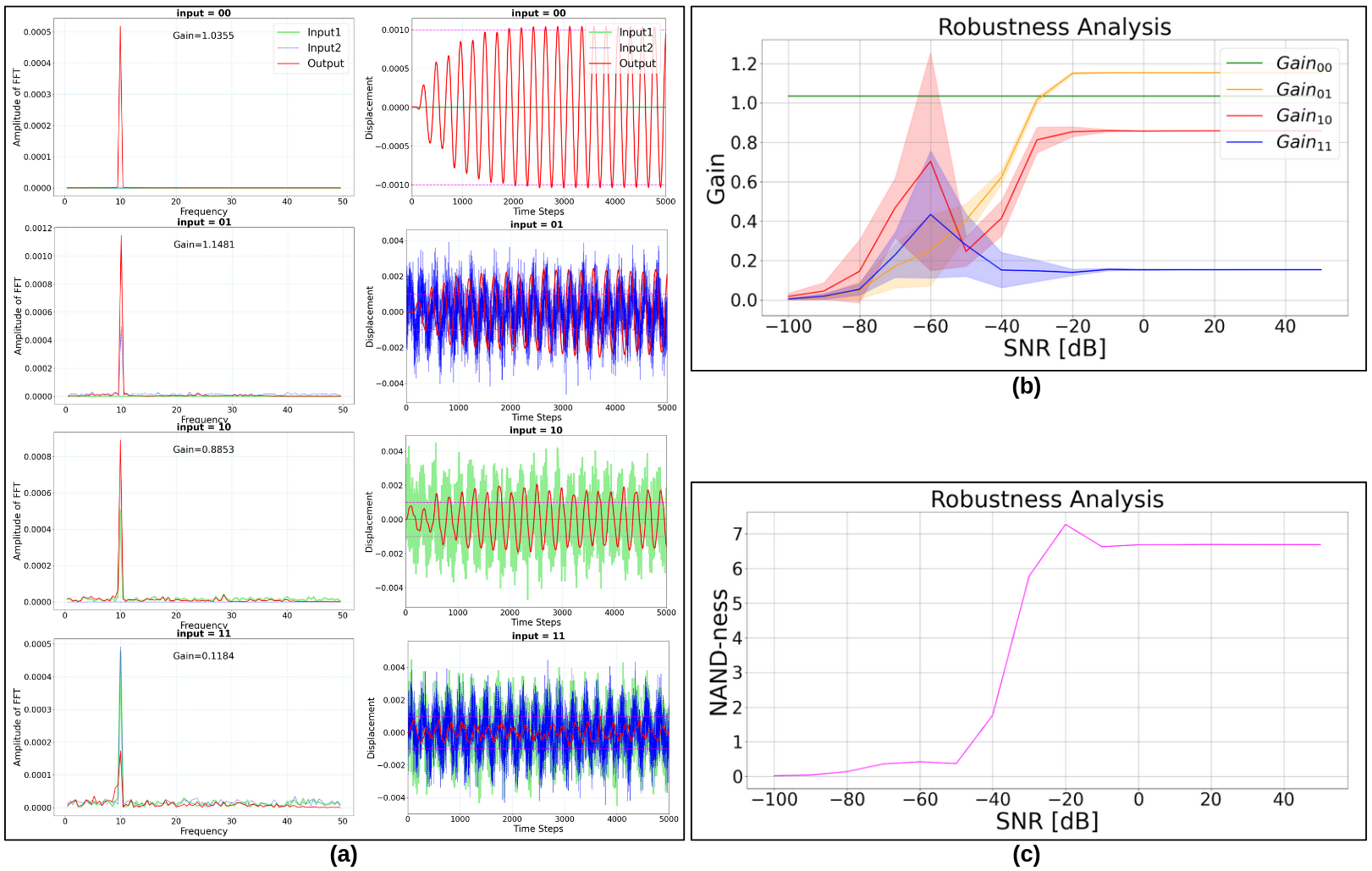}
\caption{Robustness analysis. Gaussian white noise is added to the input signals. Panel (a) shows the functionality of an evolved material with $SNR=-20dB$. Panels (b) and (c) report effects on individual gains, and NAND-ness, for different SNRs. The plots are averaged over $20$ random trials.}\label{fig13}
\end{figure*}

Noise is an inevitable component of all signals in real-world applications. So it is critical to understand its effect on a system's performance and analyze its robustness when  subject to various degrees and types of noise. In this section, we investigated the robustness of our evolved mechanical NAND gates. We used one of the evolved particle configurations from Section 2 (Fig.\ref{fig5}) as a case study. We added increasing amounts of Additive White Gaussian Noise (AWGN) to the input signals and recorded the system's response at each signal-to-noise ratio (SNR) tested. Fig.~\ref{fig13}b,c report the effects on the NAND-ness (Eqn.~\ref{eq3}) and individual gain values for that material.

Although the added noise contains all of the frequency components (see the frequency spectrum in Fig.\ref{fig13}a), we observe that if the SNR exceeds $-20dB$, the noise does not interfere with the NAND operation in the material. 

\section{Conclusion and Future Work}
In this paper, we utilized evolutionary algorithms to design granular materials capable of polycomputation. Our granular materials here were simulated with an overly simplified model which assumed frictionless circular disks. Additionally, the only parameters considered in the optimization were the stiffness of the particles and the placement of the input ports. There are various other parameters such as the particle shapes, sizes, masses, and nonlinear inter-particle contacts that can be considered in building a more realistic system with possibly more computational power. We will pursue this possibility in our future research. We also plan to study the nonlinear dynamics of the evolved solutions and develop a formal analysis to investigate the reasoning behind the stiffness patterns that enable the embedding of the logical functions at various input frequencies.

In Section 3 of the paper, we computed the amount of NAND functionality at the particles other than the pre-determined output port and realized that other particles are found that also possess NAND functionality. This might suggest the distributed nature of computation in the material and provide opportunities to design more computationally dense materials. In this regard, the information-theoretic measures can be used to provide more insight into the flow of information processing in the material and shed light on some of these observed phenomena.

Realizing physical polycomputational materials remains a future challenge. In preparation for this, we are developing granular assemblies whose configurations may be evolved to transform them into computational materials. Acoustic granular switches have been shown in one-dimensional granular systems, but such systems have primarily used passive particles\cite{Feng14}. Recent advances in material science have led to variable stiffness composites made out of phase-changing materials, and we have used such materials to develop variable stiffness particles.

\begin{figure}[h!]
    \centering
    \includegraphics[scale=0.2]{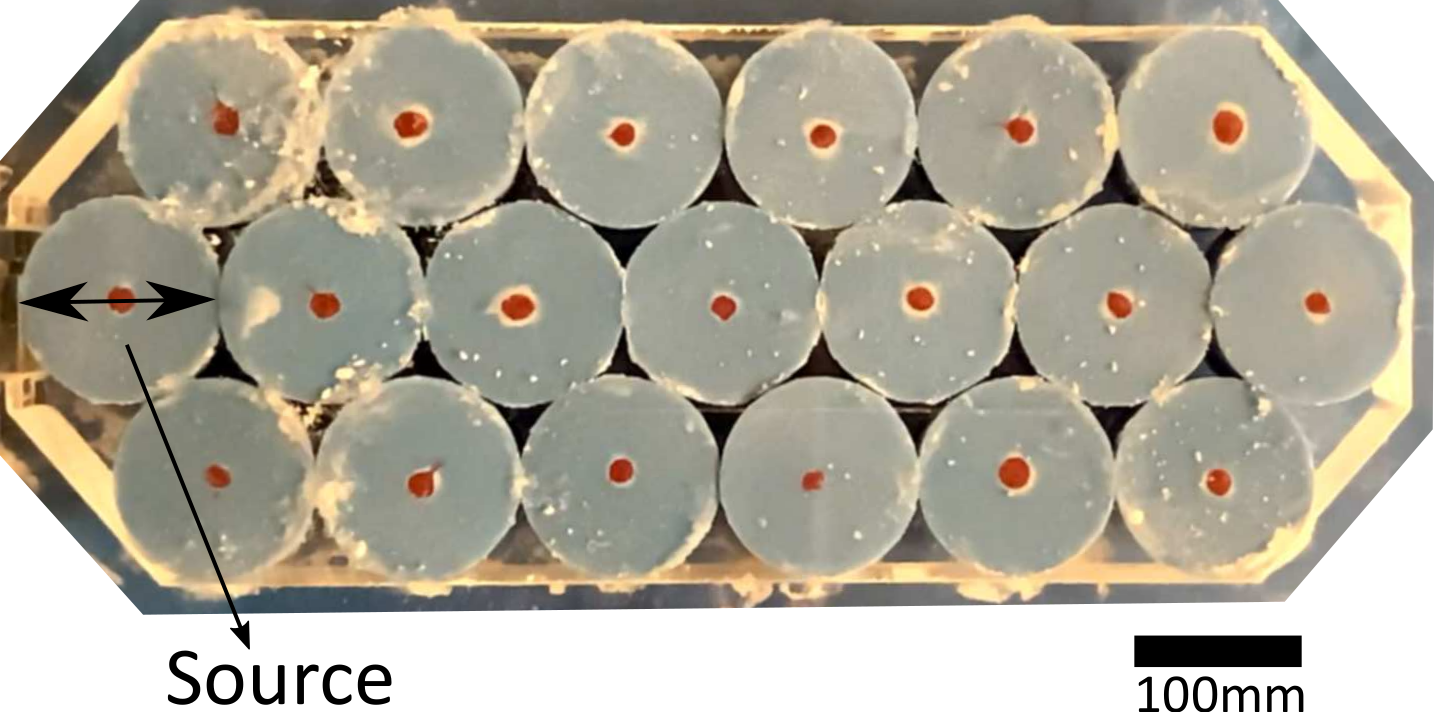}
    \caption{Two-dimensional granular system in hardware. Each circle corresponds to a particle made of silicone elastomers. Oscillating the source particle at a fixed frequency can create vibrations that correspond to the eigenmodes of the system.}\label{fig16}
\end{figure}

Our particles encapsulate a low melting-point alloy (Field's metal) in a soft silicone elastomer with a heater embedded inside. By locally heating individual particles, we can change the stiffness of each particle. Fig.~\ref{fig16} shows a two-dimensional granular system where each particle is made out of a soft silicone elastomer. By sinusoidally actuating the source particle, we can generate a sustained normal mode in the system. Using this class of polymeric materials can allow us to create variable stiffness granular materials that can be used to perform complex computations.

\begin{acks}
We would like to thank Dong Wang for his help in developing the simulator for the granular system used in the experiments in this paper. We would like to acknowledge financial support from the National Science Foundation under the DMREF program (award number: $2118810$). Computations were performed, in part, on the Vermont Advanced Computing Center.
\end{acks}

\bibliographystyle{ACM-Reference-Format}
\bibliography{main}


\begin{thebibliography}{24}


\ifx \showCODEN    \undefined \def \showCODEN     #1{\unskip}     \fi
\ifx \showDOI      \undefined \def \showDOI       #1{#1}\fi
\ifx \showISBNx    \undefined \def \showISBNx     #1{\unskip}     \fi
\ifx \showISBNxiii \undefined \def \showISBNxiii  #1{\unskip}     \fi
\ifx \showISSN     \undefined \def \showISSN      #1{\unskip}     \fi
\ifx \showLCCN     \undefined \def \showLCCN      #1{\unskip}     \fi
\ifx \shownote     \undefined \def \shownote      #1{#1}          \fi
\ifx \showarticletitle \undefined \def \showarticletitle #1{#1}   \fi
\ifx \showURL      \undefined \def \showURL       {\relax}        \fi
\providecommand\bibfield[2]{#2}
\providecommand\bibinfo[2]{#2}
\providecommand\natexlab[1]{#1}
\providecommand\showeprint[2][]{arXiv:#2}

\bibitem[Amos et~al\mbox{.}(2002)]%
        {amos2002topics}
\bibfield{author}{\bibinfo{person}{Martyn Amos}, \bibinfo{person}{Gheorghe
  P{\u{a}}un}, \bibinfo{person}{Grzegorz Rozenberg}, {and}
  \bibinfo{person}{Arto Salomaa}.} \bibinfo{year}{2002}\natexlab{}.
\newblock \showarticletitle{Topics in the theory of DNA computing}.
\newblock \bibinfo{journal}{\emph{Theoretical computer science}}
  \bibinfo{volume}{287}, \bibinfo{number}{1} (\bibinfo{year}{2002}),
  \bibinfo{pages}{3--38}.
\newblock


\bibitem[Bilal et~al\mbox{.}(2017)]%
        {bilal2017bistable}
\bibfield{author}{\bibinfo{person}{Osama~R Bilal}, \bibinfo{person}{Andr{\'e}
  Foehr}, {and} \bibinfo{person}{Chiara Daraio}.}
  \bibinfo{year}{2017}\natexlab{}.
\newblock \showarticletitle{Bistable metamaterial for switching and cascading
  elastic vibrations}.
\newblock \bibinfo{journal}{\emph{Proceedings of the National Academy of
  Sciences}} \bibinfo{volume}{114}, \bibinfo{number}{18}
  (\bibinfo{year}{2017}), \bibinfo{pages}{4603--4606}.
\newblock


\bibitem[Bitzek et~al\mbox{.}(2006)]%
        {bitzek2006structural}
\bibfield{author}{\bibinfo{person}{Erik Bitzek}, \bibinfo{person}{Pekka
  Koskinen}, \bibinfo{person}{Franz G{\"a}hler}, \bibinfo{person}{Michael
  Moseler}, {and} \bibinfo{person}{Peter Gumbsch}.}
  \bibinfo{year}{2006}\natexlab{}.
\newblock \showarticletitle{Structural relaxation made simple}.
\newblock \bibinfo{journal}{\emph{Physical review letters}}
  \bibinfo{volume}{97}, \bibinfo{number}{17} (\bibinfo{year}{2006}),
  \bibinfo{pages}{170201}.
\newblock


\bibitem[Bongard and Levin(2022)]%
        {bongard2022there}
\bibfield{author}{\bibinfo{person}{Joshua Bongard} {and}
  \bibinfo{person}{Michael Levin}.} \bibinfo{year}{2022}\natexlab{}.
\newblock \showarticletitle{There's Plenty of Room Right Here: Biological
  Systems as Evolved, Overloaded, Multi-scale Machines}.
\newblock \bibinfo{journal}{\emph{arXiv preprint arXiv:2212.10675}}
  (\bibinfo{year}{2022}).
\newblock


\bibitem[El~Helou et~al\mbox{.}(2022)]%
        {el2022mechanical}
\bibfield{author}{\bibinfo{person}{Charles El~Helou}, \bibinfo{person}{Benjamin
  Grossmann}, \bibinfo{person}{Christopher~E Tabor}, \bibinfo{person}{Philip~R
  Buskohl}, {and} \bibinfo{person}{Ryan~L Harne}.}
  \bibinfo{year}{2022}\natexlab{}.
\newblock \showarticletitle{Mechanical integrated circuit materials}.
\newblock \bibinfo{journal}{\emph{Nature}} \bibinfo{volume}{608},
  \bibinfo{number}{7924} (\bibinfo{year}{2022}), \bibinfo{pages}{699--703}.
\newblock


\bibitem[Hey(1999)]%
        {hey1999quantum}
\bibfield{author}{\bibinfo{person}{Tony Hey}.} \bibinfo{year}{1999}\natexlab{}.
\newblock \showarticletitle{Quantum computing: an introduction}.
\newblock \bibinfo{journal}{\emph{Computing \& Control Engineering Journal}}
  \bibinfo{volume}{10}, \bibinfo{number}{3} (\bibinfo{year}{1999}),
  \bibinfo{pages}{105--112}.
\newblock


\bibitem[Hooker(2021)]%
        {hooker2021hardware}
\bibfield{author}{\bibinfo{person}{Sara Hooker}.}
  \bibinfo{year}{2021}\natexlab{}.
\newblock \showarticletitle{The hardware lottery}.
\newblock \bibinfo{journal}{\emph{Commun. ACM}} \bibinfo{volume}{64},
  \bibinfo{number}{12} (\bibinfo{year}{2021}), \bibinfo{pages}{58--65}.
\newblock


\bibitem[Ion et~al\mbox{.}(2017)]%
        {ion2017digital}
\bibfield{author}{\bibinfo{person}{Alexandra Ion}, \bibinfo{person}{Ludwig
  Wall}, \bibinfo{person}{Robert Kovacs}, {and} \bibinfo{person}{Patrick
  Baudisch}.} \bibinfo{year}{2017}\natexlab{}.
\newblock \showarticletitle{Digital mechanical metamaterials}. In
  \bibinfo{booktitle}{\emph{Proceedings of the 2017 CHI Conference on Human
  Factors in Computing Systems}}. \bibinfo{pages}{977--988}.
\newblock


\bibitem[Kaspar et~al\mbox{.}(2021)]%
        {kaspar2021rise}
\bibfield{author}{\bibinfo{person}{C Kaspar}, \bibinfo{person}{BJ Ravoo},
  \bibinfo{person}{Wilfred~G van~der Wiel}, \bibinfo{person}{SV Wegner}, {and}
  \bibinfo{person}{WHP Pernice}.} \bibinfo{year}{2021}\natexlab{}.
\newblock \showarticletitle{The rise of intelligent matter}.
\newblock \bibinfo{journal}{\emph{Nature}} \bibinfo{volume}{594},
  \bibinfo{number}{7863} (\bibinfo{year}{2021}), \bibinfo{pages}{345--355}.
\newblock


\bibitem[Kim and Yang(2019)]%
        {kim2019wave}
\bibfield{author}{\bibinfo{person}{Eunho Kim} {and} \bibinfo{person}{Jinkyu
  Yang}.} \bibinfo{year}{2019}\natexlab{}.
\newblock \showarticletitle{Wave propagation in granular metamaterials}.
\newblock \bibinfo{journal}{\emph{Functional Composites and Structures}}
  \bibinfo{volume}{1}, \bibinfo{number}{1} (\bibinfo{year}{2019}),
  \bibinfo{pages}{012002}.
\newblock


\bibitem[Li et~al\mbox{.}(2014)]%
        {Feng14}
\bibfield{author}{\bibinfo{person}{Feng Li}, \bibinfo{person}{Paul Anzel},
  \bibinfo{person}{Jinkyu Yang}, \bibinfo{person}{Panayotis Kevrekidis}, {and}
  \bibinfo{person}{Chiara Daraio}.} \bibinfo{year}{2014}\natexlab{}.
\newblock \showarticletitle{Granular acoustic switches and logic elements}.
\newblock \bibinfo{journal}{\emph{Nature communications}}  \bibinfo{volume}{5}
  (\bibinfo{date}{10} \bibinfo{year}{2014}), \bibinfo{pages}{5311}.
\newblock
\urldef\tempurl%
\url{https://doi.org/10.1038/ncomms6311}
\showDOI{\tempurl}


\bibitem[Markovi{\'c} et~al\mbox{.}(2020)]%
        {markovic2020physics}
\bibfield{author}{\bibinfo{person}{Danijela Markovi{\'c}},
  \bibinfo{person}{Alice Mizrahi}, \bibinfo{person}{Damien Querlioz}, {and}
  \bibinfo{person}{Julie Grollier}.} \bibinfo{year}{2020}\natexlab{}.
\newblock \showarticletitle{Physics for neuromorphic computing}.
\newblock \bibinfo{journal}{\emph{Nature Reviews Physics}} \bibinfo{volume}{2},
  \bibinfo{number}{9} (\bibinfo{year}{2020}), \bibinfo{pages}{499--510}.
\newblock


\bibitem[Miskin and Jaeger(2013)]%
        {miskin2013adapting}
\bibfield{author}{\bibinfo{person}{Marc~Z Miskin} {and}
  \bibinfo{person}{Heinrich~M Jaeger}.} \bibinfo{year}{2013}\natexlab{}.
\newblock \showarticletitle{Adapting granular materials through artificial
  evolution}.
\newblock \bibinfo{journal}{\emph{Nature materials}} \bibinfo{volume}{12},
  \bibinfo{number}{4} (\bibinfo{year}{2013}), \bibinfo{pages}{326--331}.
\newblock


\bibitem[Nakajima(2020)]%
        {nakajima2020physical}
\bibfield{author}{\bibinfo{person}{Kohei Nakajima}.}
  \bibinfo{year}{2020}\natexlab{}.
\newblock \showarticletitle{Physical reservoir computing—an introductory
  perspective}.
\newblock \bibinfo{journal}{\emph{Japanese Journal of Applied Physics}}
  \bibinfo{volume}{59}, \bibinfo{number}{6} (\bibinfo{year}{2020}),
  \bibinfo{pages}{060501}.
\newblock


\bibitem[Parsa et~al\mbox{.}(2022a)]%
        {parsa2022evolving}
\bibfield{author}{\bibinfo{person}{Atoosa Parsa}, \bibinfo{person}{Dong Wang},
  \bibinfo{person}{Corey~S O'Hern}, \bibinfo{person}{Mark~D Shattuck},
  \bibinfo{person}{Rebecca Kramer-Bottiglio}, {and} \bibinfo{person}{Josh
  Bongard}.} \bibinfo{year}{2022}\natexlab{a}.
\newblock \showarticletitle{Evolving programmable computational metamaterials}.
  In \bibinfo{booktitle}{\emph{Proceedings of the Genetic and Evolutionary
  Computation Conference}}. \bibinfo{pages}{122--129}.
\newblock


\bibitem[Parsa et~al\mbox{.}(2022b)]%
        {ourPaper}
\bibfield{author}{\bibinfo{person}{Atoosa Parsa}, \bibinfo{person}{Dong Wang},
  \bibinfo{person}{Corey~S. O’Hern}, \bibinfo{person}{Mark~D. Shattuck},
  \bibinfo{person}{Rebecca Kramer-Bottiglio}, {and} \bibinfo{person}{Josh
  Bongard}.} \bibinfo{year}{2022}\natexlab{b}.
\newblock \showarticletitle{Evolution of Acoustic Logic Gates in Granular
  Metamaterials}. In \bibinfo{booktitle}{\emph{International Conference on the
  Applications of Evolutionary Computation (Part of EvoStar)}}. Springer.
\newblock
\newblock
\shownote{(in press)}.


\bibitem[Rajappan et~al\mbox{.}(2022)]%
        {rajappan2022logic}
\bibfield{author}{\bibinfo{person}{Anoop Rajappan}, \bibinfo{person}{Barclay
  Jumet}, \bibinfo{person}{Rachel~A Shveda}, \bibinfo{person}{Colter~J Decker},
  \bibinfo{person}{Zhen Liu}, \bibinfo{person}{Te~Faye Yap},
  \bibinfo{person}{Vanessa Sanchez}, {and} \bibinfo{person}{Daniel~J Preston}.}
  \bibinfo{year}{2022}\natexlab{}.
\newblock \showarticletitle{Logic-enabled textiles}.
\newblock \bibinfo{journal}{\emph{Proceedings of the National Academy of
  Sciences}} \bibinfo{volume}{119}, \bibinfo{number}{35}
  (\bibinfo{year}{2022}), \bibinfo{pages}{e2202118119}.
\newblock


\bibitem[Schmidt and Lipson(2010)]%
        {schmidt2010age}
\bibfield{author}{\bibinfo{person}{Michael~D Schmidt} {and}
  \bibinfo{person}{Hod Lipson}.} \bibinfo{year}{2010}\natexlab{}.
\newblock \showarticletitle{Age-fitness pareto optimization}. In
  \bibinfo{booktitle}{\emph{Proceedings of the 12th annual conference on
  Genetic and evolutionary computation}}. \bibinfo{pages}{543--544}.
\newblock


\bibitem[Serra-Garcia(2019)]%
        {serra2019turing}
\bibfield{author}{\bibinfo{person}{Marc Serra-Garcia}.}
  \bibinfo{year}{2019}\natexlab{}.
\newblock \showarticletitle{Turing-complete mechanical processor via automated
  nonlinear system design}.
\newblock \bibinfo{journal}{\emph{Physical Review E}} \bibinfo{volume}{100},
  \bibinfo{number}{4} (\bibinfo{year}{2019}), \bibinfo{pages}{042202}.
\newblock


\bibitem[Silva et~al\mbox{.}(2014)]%
        {silva2014performing}
\bibfield{author}{\bibinfo{person}{Alexandre Silva}, \bibinfo{person}{Francesco
  Monticone}, \bibinfo{person}{Giuseppe Castaldi}, \bibinfo{person}{Vincenzo
  Galdi}, \bibinfo{person}{Andrea Al{\`u}}, {and} \bibinfo{person}{Nader
  Engheta}.} \bibinfo{year}{2014}\natexlab{}.
\newblock \showarticletitle{Performing mathematical operations with
  metamaterials}.
\newblock \bibinfo{journal}{\emph{Science}} \bibinfo{volume}{343},
  \bibinfo{number}{6167} (\bibinfo{year}{2014}), \bibinfo{pages}{160--163}.
\newblock


\bibitem[Solli and Jalali(2015)]%
        {solli2015analog}
\bibfield{author}{\bibinfo{person}{Daniel~R Solli} {and}
  \bibinfo{person}{Bahram Jalali}.} \bibinfo{year}{2015}\natexlab{}.
\newblock \showarticletitle{Analog optical computing}.
\newblock \bibinfo{journal}{\emph{Nature Photonics}} \bibinfo{volume}{9},
  \bibinfo{number}{11} (\bibinfo{year}{2015}), \bibinfo{pages}{704--706}.
\newblock


\bibitem[Treml et~al\mbox{.}(2018)]%
        {treml2018origami}
\bibfield{author}{\bibinfo{person}{Benjamin Treml}, \bibinfo{person}{Andrew
  Gillman}, \bibinfo{person}{Philip Buskohl}, {and} \bibinfo{person}{Richard
  Vaia}.} \bibinfo{year}{2018}\natexlab{}.
\newblock \showarticletitle{Origami mechanologic}.
\newblock \bibinfo{journal}{\emph{Proceedings of the National Academy of
  Sciences}} \bibinfo{volume}{115}, \bibinfo{number}{27}
  (\bibinfo{year}{2018}), \bibinfo{pages}{6916--6921}.
\newblock


\bibitem[Wu et~al\mbox{.}(2019)]%
        {wu2019active}
\bibfield{author}{\bibinfo{person}{Qikai Wu}, \bibinfo{person}{Chunyang Cui},
  \bibinfo{person}{Thibault Bertrand}, \bibinfo{person}{Mark~D Shattuck}, {and}
  \bibinfo{person}{Corey~S O'Hern}.} \bibinfo{year}{2019}\natexlab{}.
\newblock \showarticletitle{Active acoustic switches using two-dimensional
  granular crystals}.
\newblock \bibinfo{journal}{\emph{Physical Review E}} \bibinfo{volume}{99},
  \bibinfo{number}{6} (\bibinfo{year}{2019}), \bibinfo{pages}{062901}.
\newblock


\bibitem[Yasuda et~al\mbox{.}(2021)]%
        {yasuda2021mechanical}
\bibfield{author}{\bibinfo{person}{Hiromi Yasuda}, \bibinfo{person}{Philip~R
  Buskohl}, \bibinfo{person}{Andrew Gillman}, \bibinfo{person}{Todd~D Murphey},
  \bibinfo{person}{Susan Stepney}, \bibinfo{person}{Richard~A Vaia}, {and}
  \bibinfo{person}{Jordan~R Raney}.} \bibinfo{year}{2021}\natexlab{}.
\newblock \showarticletitle{Mechanical computing}.
\newblock \bibinfo{journal}{\emph{Nature}} \bibinfo{volume}{598},
  \bibinfo{number}{7879} (\bibinfo{year}{2021}), \bibinfo{pages}{39--48}.
\newblock


\end{thebibliography}

\end{document}